# An Indoor Fingerprinting Localization Approach for ZigBee Wireless Sensor Networks

**Tareq Alhmiedat**
*Computer Science Department, Zarqa University, Zarqa Street, Jordan*
E-mail: t.alhmiedat@zu.edu.jo

**Ghassan Samara**
*Computer Science Department, Zarqa University, Zarqa Street, Jordan*
E-mail: gsamara@zu.edu.jo

**Amer O. Abu Salem**
*Computer Information Systems Department, Zarqa University, Zarqa, Jordan*
E-mail: abusalem@zu.edu.jo

## Abstract

Location tracking systems are increasingly becoming the focus of research in the field of Wireless Sensor Network (WSN). Received Signal Strength (RSS)-based localization systems are at the forefront of tracking research applications. Radio location fingerprinting is one of the most promising indoor positioning approaches due to its powerful in terms of accuracy and cost. However, fingerprinting systems require the collection of a large number of reference points in the tracking area to achieve reasonable localization accuracy. In this paper, we propose a fingerprinting localization approach based on a RSS technique. The proposed system does not require gathering a large number of reference points and offers good localization accuracy indoors. The implemented approach is based on dividing the tracking area into subareas and assigning a unique feature to each subarea through ranging the RSS values from different reference points. In order to test the proposed system's efficiency, a number of real experiments have been conducted using Jennic sensor nodes.

## 1. Introduction

Wireless Sensor Network (WSN) has become a vital research area, due to their wide ranging applications including civilian, industrial, agricultural, and military [1, 2, 3]. A sensor network consists of sensor nodes which are small in size, low in cost, and have short communication range. Usually, a sensor device consists of four main subsystems (computing, sensing, communication, and power supply subsystems).

Researchers have focused on different aspects of WSN, such as hardware design, routing, security, and localization [4, 5, 6]. One of the critical aspects which needs to be taken into consideration is localizing mobile targets through distributed sensor networks. Node localization is the problem of finding the geographical location of a mobile target node (the node with unknown location), based on other beacon nodes (nodes with fixed known locations).

Target tracking applications using WSNs have received much attention recently. This attention focuses on the need to achieve high localization accuracy, without incurring a large cost, form factor and power-consumption per node. There are many localization and tracking systems have been designed and implemented recently, summarized in [7, 8, 9]. A well known localization approach is the Received Signal Strength (RSS). RSS-based localization systems are one of the most popular and cheap techniques and are increasingly accepted as a positioning solution for localizing target nodes in both indoor and outdoor environments. RSS-based localization systems work by converting the Signal Strength (SS) to a transmitter-receiver using separate distance measurements. However, RSS values can be affected by walls and obstacles which may reflect and propagate the signals, therefore offering a non-linear transformation between the RSS values and the location. Due to the aforementioned limitations, deploying RSS-based localization systems in indoor environments becomes a complicated task which is difficult to engineer using classical mathematical models.

The RSS information can be used to estimate the distance between the transmitter and the receiver in two ways: the first uses the signal propagation model to convert SS to a distance measurement, using previous knowledge about the beacon nodes' locations, and deploys a geometry method to compute the location of target nodes. This is known as a triangulation localization method [10]. The second approach is based on the behavior of signal propagation and information about the geometry of the building to convert RSS values into distance values; this is known as a fingerprinting localization method.

The complexity of applying a triangulation approach arises from the need to accurately obtain the distance measurements from the RSS as indoor radio signal propagation is very complicated because of signal attenuation due to distance, the effect of multipath propagation, and penetration losses through obstacles.

On the other hand, fingerprinting systems require only the collection of RSS values at several locations to form a database of location fingerprints. Deployment of the fingerprinting-based localization system is usually divided into two main phases: 1) Offline phase: this includes measuring the location of a mobile target in several coordinates and storing the collected RSS values at each point with the corresponding location in a database file; 2) Online phase: the mobile target collects several RSS values from different beacon nodes in its range and sends the data to a server which applies a positioning algorithm to estimate the mobile target's location [11].

There are two main challenges to designing and developing a fingerprinting system. First, there is the problem of collecting the RSS samples and storing them in a database file; this process requires a long period of time, particularly when the localization system is deployed in a large area. Second, the searching procedure through the stored samples is time consuming. In this paper, we propose a fingerprinting-based localization approach which aims to reduce the total number of reference points that need to be collected in the offline phase while achieving low localization error of between 1 and 3.5 meters. Moreover, the proposed system assigns a unique feature to each small area in order to reduce the time needed to carry out the searching procedure.

The main contributions of this paper are as follows: a) unlike the existing approaches which mainly focus on simulation, a new fingerprinting localization approach is introduced practically using ZigBee sensor nodes; b) it is demonstrated analytically that the proposed localization approach requires a lower number of reference points to be gathered than the existing neural network based approaches; c) through experiments, it is shown that the proposed localization method achieves a low level of localization error in complex environments.

This paper is organized as follows: In Section 2, the related systems proposed recently are introduced. In Section 3, a new fingerprinting localization approach is presented, while Section 4 reviews the implementation of the proposed system on real ZigBee-based sensor devices. Section 5 offers the evaluation of the proposed fingerprinting localization system and a discussion is presented in Section 6. Finally, Section 7 draws conclusions regarding the proposed system and presents suggestions for future work.

## 2. Related Work

Even though WSNs were not designed and developed for the purpose of localization applications, measuring the RSS values for each transmitted message could offer localization information of mobile targets. Fingerprinting localization systems are relatively simple and cheap compared to other methods such as angle-of-arrival (AOA), time-of-arrival (TOA), and time difference of arrival (TDOA) systems, as these systems require attaching additional sensor device to each beacon node. In this section, the fingerprinting-based localization approaches are categorized into two categories based on the localization method used in the estimation phase: database-based and neural network-based methods.

First, the database-based approaches are considered. The systems proposed in [12, 13] include a WLAN-based indoor localization system based on fingerprinting method. The developed approaches are based on collecting RSS values from several reference points distributed over the tracking area and storing them in a database file. The mobile target estimates its location by comparing the RSS values collected from the beacon nodes and the RSS values stored in the database.

RADAR is a well-known tracking system proposed in [14], which operates by recording and processing information regarding the signal strength at multiple base stations in order to provide overlapping coverage of the area of interest. The work presented in [15] includes a research and a development of a fingerprinting-indoor positioning system using the received signal strength, which its based on two different protocol stacks: BitCloud and OpenMac.

Cortina is a distributed real time location system designed to track people indoors based on IEEE802.15.4 radio standard and RFID technology. Cortina avoids the need for manual calibration by adopting a collaborative approach that uses RSS measurements collected by the fixed nodes [16]. A hybrid localization approach using WLAN, which consists of two main stages: first, a fingerprinting method, with a fast training phase in order to obtain the location for the mobile target, indicates which room the mobile target is located in. Second, a triangulation is used to compute the mobile target location precisely. It was shown that the proposed hybrid system is more accurate than the triangulation approach. However, the proposed hybrid system offers lower localization accuracy than the fingerprinting method [17]. In [18], a hybrid localization approach based on WSN was proposed, that integrates two techniques: RF mapping and cooperative ranging to enhance the localization accuracy.

In order to reduce the time required for the searching process in the online phase. The work proposed in [19] includes a new search strategy for radio fingerprint matching method, which significantly reduces the search operations with a little effect on the localization accuracy.

The second approach is neural network-based. The work presented in [20] includes an indoor localization system based on a modular multi-layer perceptron, where three neural network modules were designed in order to cover the absence of signals from access points. The proposed work in [21] includes a fingerprinting-based localization system with a localization error of 3 meters, where a total number of 5 reference points were needed for the training process in the offline phase. In [22], the authors compared the performance of three different families of neural networks for localization applications using WSN: Multi-Layer Perceptron (MLP), Radial Basis Function (RBF), and Recurrent Neural Networks (RNN). Through experiments, RBF module was the best choice as it offered high localization accuracy, and minimal computational and memory requirements.

## 3. Radio Frequency Fingerprinting Approach

The indoor fingerprinting positioning approach developed in this paper consists of three main phases: the Creation of the fingerprint table phase, the Feature identification phase, and the Estimation phase. The first two phases are carried out during the offline stage while the third one is carried out in the online stage. Figure 1 depicts the proposed approach and Table 1 includes the definitions of several parameters.

**Figure 1:** RF Fingerprinting approach

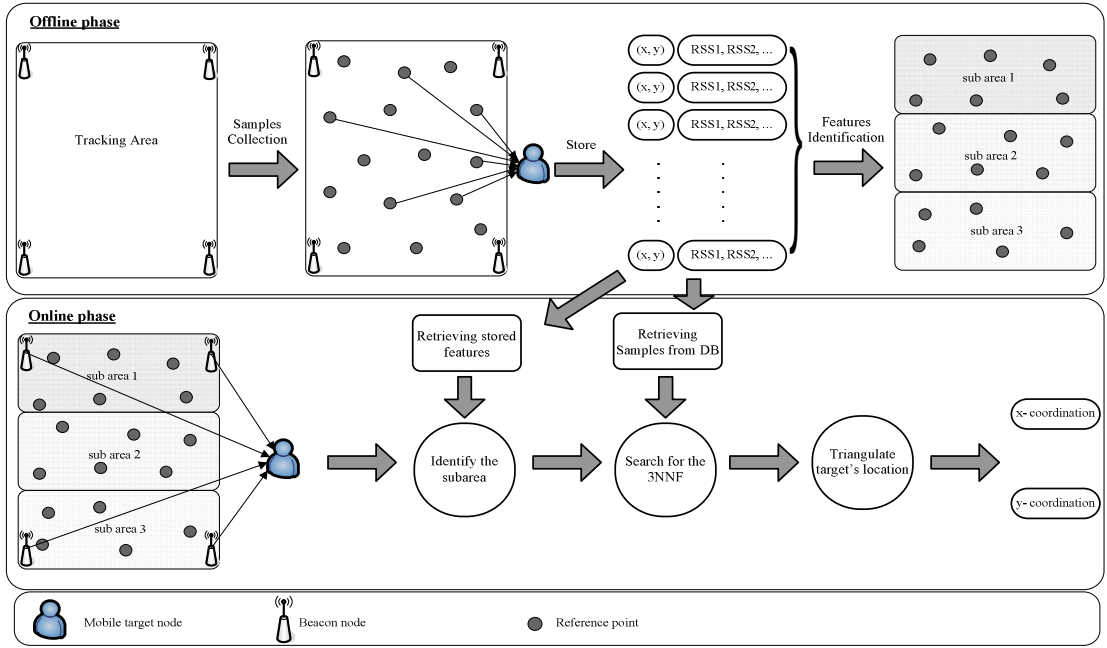

**Table 1:**    A Sample of Companies Listed in Tehran Stock Exchange

| Name | Parameter | Definition |
|------|-----------|------------|
| Beacon node | $b$ | A node with fixed position and known coordinates *(x, y)* |
| Reference point | $r$ | A point with known RSS values collected previously from beacon nodes, and identified *(x, y)* coordinates |
| Estimated point | $e$ | A point which needs to be found based on the RSS values collected from beacon nodes |
| Criteria | $C_{Ai}$ | A set of features distinguishes subarea $A_i$ from other subareas |

## 3.1. Creation of Fingerprint Table Phase

Assume the tracking area is divided into grid points, and the coordinate for each grid point is $P_i = (x_i, y_i)$, $i = 1, 2, ..., n$ where $n$ is the total number of grid points. Also $m$ is the total number of reference points, where $m < n$. This phase includes two sub-stages:

1. Collect RSS values from several beacon nodes $\{b_1, b_2, ..., b_z\}$ at each reference point $r_k$ and store them in a database file. The collected reference points have to be spread as evenly as possible in the tracking area of interest.

2. Divide the tracking area of interest into subareas as $\{A_1, A_2, ..., A_u\}$, where $u$ is the total number of subareas.

   In the first stage, the mobile target $t$ goes through a total number of reference grid points $m$. The mobile target $t$ starts to receive location messages from beacon nodes at each reference point $r_k = (x_k, y_k)$. Let $rss_{b_j}^k$ refers to the average RSS values from the $b_j$ th beacon node at $r_k$ reference point. We can establish a RSS vector at each reference grid point as $S_k = \left\{ rss_{b_1}^k, rss_{b_2}^k, rss_{b_3}^k, ..., rss_{b_z}^k \right\}$, where $z$ is the total number of beacon nodes.

   The second stage involves dividing the tracking area into subareas based on assigning a unique feature to each subarea. There are three reasons behind dividing the tracking area into subareas. First, it sometimes enhances the localization accuracy, by searching the reference points which set in the same subarea where the mobile target settles in. Second, any changes in the environment after the collection phase can be recovered by recollecting reference points from that subarea. And third, dividing the search area into smaller subareas drastically reduces the search process and space. Based on the previous location of a mobile target, only the last known subarea and the immediately surrounding reference points need to be searched to find the new location of the mobile target.

A labeled training set (including input signals, corresponding output locations, subareas and features) is stored in the database, as shown in Table 2.

**Table 2:**     Structure of database

| Vector | RSS values | Corresponding subarea | Feature | Location |
|--------|-----------|----------------------|---------|----------|
| $S_1$ | $\left\{ rss_{b_1}^1, rss_{b_2}^1, rss_{b_3}^1, ...., rss_{b_z}^1 \right\}$ | $A_1$ | $C_{A1}$ | $(x_1, y_1)$ |
| $S_2$ | $\left\{ rss_{b_1}^2, rss_{b_2}^2, rss_{b_3}^2, ...., rss_{b_z}^2 \right\}$ | $A_2$ | $C_{A2}$ | $(x_2, y_2)$ |
| $S_n$ | $\left\{ rss_{b_1}^n, rss_{b_2}^n, rss_{b_3}^n, ...., rss_{b_z}^n \right\}$ | $A_l$ | $C_{Al}$ | $(x_s, y_s)$ |

### 3.2. Feature Identification Phase

In this phase, a unique feature is identified for each single subarea. The features can be identified based on the RSS values collected from adjacent beacon nodes $\{b_1, b_2, ..., b_z\}$ at the reference points in the subarea $A_i$. Assume that the set $B$ represents the beacon nodes, such that $B = \{b_1, b_2, ..., b_z\}$, and $C_{A_i}$ is a set of categories for the subarea $A_i$ such that $C_{A_i} = \left\{ Ra_{b_1}^1, Ra_{b_2}^2, ...., Ra_{b_z}^y \right\}$, where $Ra_{b_k}^i$ is the range of the RSS values for beacon node $b_k$ in the subarea $A_i$. Let $rss_{A_i}$ function as follows:

$$rss_{A_i} : R \rightarrow C_{A_i} \tag{1}$$

However, there are $N$ different subareas and consequently there are $N$ different feature functions. For instance, assume that $R = \{r_1, r_2, ..., b_s\}$ represents reference points in a given subarea $A_i$. There is a set of features $C_{A_i}$ that distinguishes the subarea $A_i$ from other subareas, where $C_{A_i} = \left\{ Ra_{b_1}, Ra_{b_2}, ...., Ra_{b_z} \right\}$, and $Ra_{b_j}$ is the range of RSS values between a beacon node $b_j$ and all the reference points $R$ in a subarea $A_i$. For instance, $Ra_{b_i} \in (62-75)$.

### 3.3. Estimation Phase

In this phase, the mobile target's location is estimated. This phase includes two main sub-stages:

1. Identify the subarea $A_i$ where the mobile target is located in, based on comparing the collected RSS values with the identified features assigned to each subarea,
2. Find out the nearest three reference points to the estimated point $e$ in the identified subarea $A_i$ based on the difference in the RSS readings in the selected subarea.

In the first stage, to locate the position of a mobile target, the RSS value is recorded between the mobile target $t$ and beacon nodes $\{b_1, b_2, ..., b_z\}$. in the mobile target's communication range. The RSS values obtained for $t$ are compared to the RSS values for known positions (reference points) stored in the database file. The subarea with the closest RSS values is assumed to be where $t$ is located.

In the second stage, to be more precise, to locate the mobile target in the subarea, the RSS values of the mobile node are compared with the features of the beacon nodes within the subarea based on Equation 2. The Three Nearest Neighbors based on Feature identification function (3NNF) is used to triangulate the mobile target's position.

$$diff\left(rss_i^e, rss_j^r\right) = \sqrt{\sum_{k=1}^{w}\left(rss_k^e - rss_k^r\right)^2} \qquad (2)$$

# 4. Experiments

In order to evaluate the proposed fingerprinting localization approach, several experiments were conducted on real sensor nodes. In this section, the main features of the proposed system including the Graphical User Interface (GUI), mobile and beacon nodes' modules, and the experimental test-beds, are illustrated.

## 4.1. Graphical User Interface (GUI)

In order to facilitate the collection and segmentation processes, the map of the tracking area of interest is stored on a laptop and a user interface based on Visual Basic.Net, was designed. In this section, the collection and segmentations processes are discussed in details.

### 4.1.1. Collection Process

In the first phase, the user has to distribute the beacon nodes over the tracking area of interest, and then collect the RSS readings at several reference points. This can be achieved by a single click on the displayed map which selects the reference point on the map and then stores the collected RSS values from each beacon node. A total number of 70 and 60 measurement points are identified and collected from distinct physical locations on test-beds 1 and 2 respectively. The graphical user interface of the proposed system is shown in Figure 2.

**Figure 2:** Graphical user interface for the proposed system

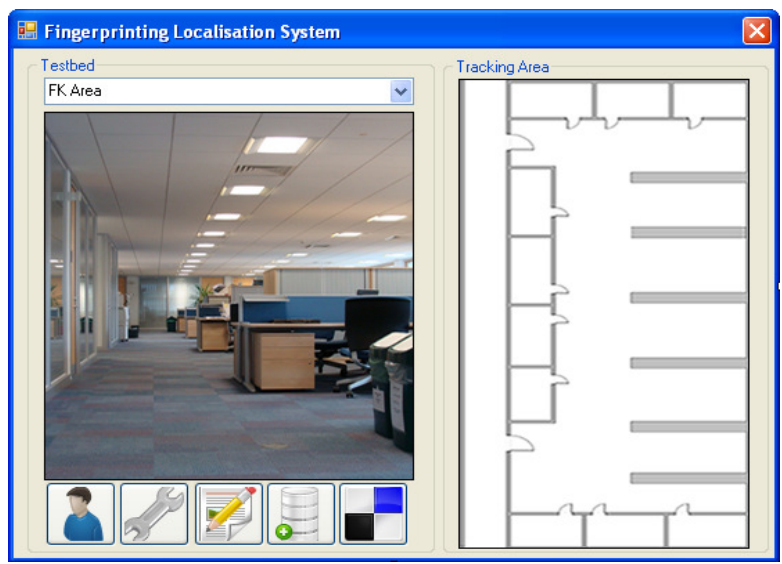

#### 4.1.2. Segmentation Process

Dividing the tracking area into subareas can be achieved in one of two ways: manual or autonomous. The manual selection can be processed by selecting the desired subarea on the displayed map. The system is designed to check whether there is a common feature among reference points, whereas, in the autonomous process, the system divides the tracking area into subareas randomly and then checks the RSS values collected from reference points in that subarea. If these reference points have a common feature, then the system will consider that area as a single subarea, and separate it from the other subareas. Otherwise, the selected subarea will be reduced and the previous step is repeated till all subareas are considered.

In the estimation phase, the mobile target collects RSS values from beacon nodes and transfers these values to the sink node; the 3NNF algorithm is applied at the server side in order to display the final position on a map.

### 4.2. Mobile and Reference Node Modules

In our experiments, we used a JN5139-EK010 sensor node platform (depicted in Figure 3) for both the mobile target and beacon nodes. This module offers low power-consumption, low processor overheads, and a low cost platform for WSNs.

**Figure 3:** JN5139 Jennic sensor module

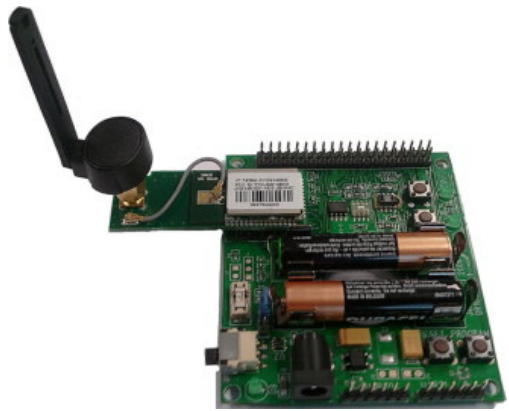

The proposed system was implemented through ZigBee network standard. ZigBee standard offers 3 main roles: coordinator, router and end-device nodes. In our experiments, the mobile target and beacon nodes were considered as router devices, while the sink node implemented the coordinator of the ZigBee network, which was responsible for collecting localization information from mobile target and transferring it to a laptop via a serial cable.

### 4.3. Experimental Test-beds

The proposed approach was tested in two different experimental test-beds in order to test its efficiency and accuracy. The experimental Test-bed 1 was located in the 1st floor of the Holywell Park building at Loughborough University. Its layout has dimensions of (41.5 × 11.3 m), as shown in Figure 4. Test-bed 1 includes obstacles and walls, which might affect the localization accuracy. The experimental Test-bed 2, depicted in Figure 5, is located in the Sir David Wallace Sports Hall at Loughborough University; it has dimensions of (30.5 × 11.3 m). For both test-beds, the origin of the coordinate system (0, 0) was placed at the left bottom corner.

**Figure 4:** Test-bed 1-FK office test environment

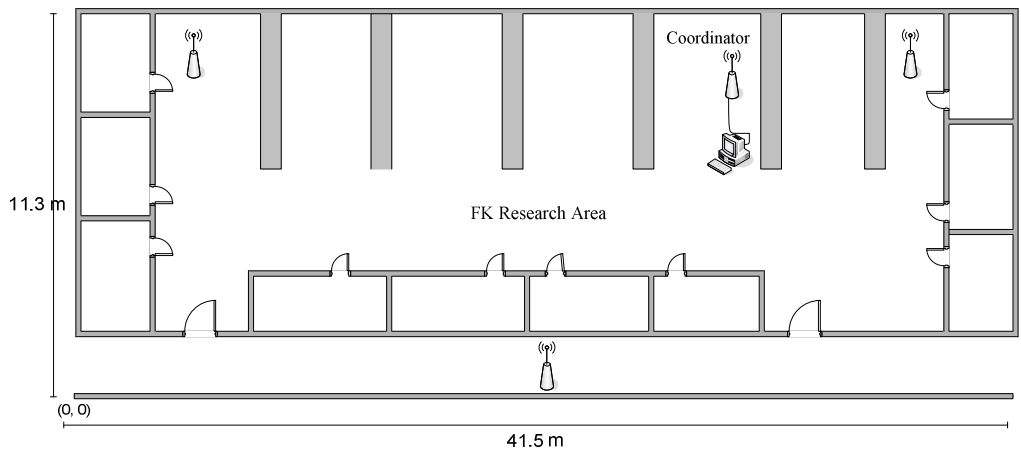

**Figure 5:** Test-bed 2-Sports hall test environment

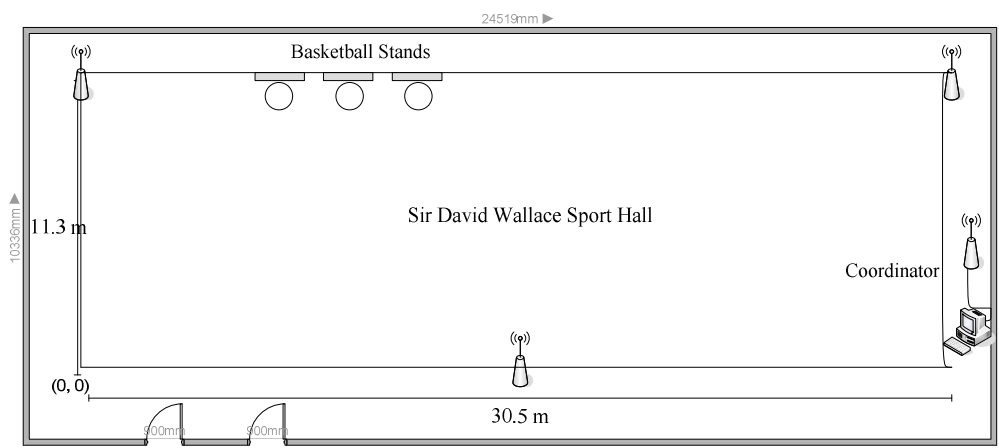

## 5. Testing Results and Performance Evaluation

To verify the validity of the proposed system described in this paper, the system was implemented through real experiments conducted in indoor environments using Jennic sensor nodes. The proposed system was evaluated through measuring the localization error, the effect of the collected number of reference points on the localization accuracy, and the efficiency of the proposed segmentation process.

### 5.1. Localization Accuracy

Localization accuracy can be described as the distance between the real and estimated locations. The localization accuracy was evaluated based on two factors: First, the effect of collected samples' number on the localization accuracy, since it was evaluated based on two different methods (3NNF localization method and the RBF neural network module). According to [22], the RBF module offers the greatest efficiency in terms of localization accuracy and memory requirements.

As shown in Figure 6, the localization accuracy for the approach proposed in this paper (3NNF) is better than the RBF neural network method as the RBF method needs to be trained at every single grid point; and this requires gathering a large number of reference points through the offline phase. Whereas the 3NNF method estimates the mobile target's position in two stages: it first finds out the subarea where the mobile target is located in, and then triangulates the target's location using the nearest three neighbor nodes.

**Figure 6:** The localization error using 3NNF method in Test-bed 1

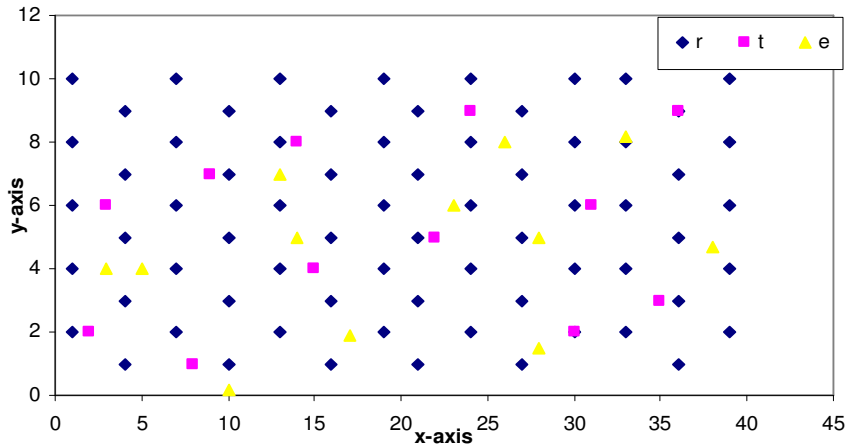

Second, the localization accuracy is evaluated based on the localization method used in the estimation phase. Figure 6 shows the localization error for the proposed fingerprinting localization method when 3NNF was used. The localization error was between 1 and 3 meters. However, the localization error was between 1.5 and 4 meters when the RBF method was used in the estimation phase, as depicted in Figure 7. In both figures, the number of reference points ( $r$ ), test points ( $t$ ), and estimated points ( $e$ ) are depicted.

**Figure 7:** The localization error using the RBF method in Test-bed 1

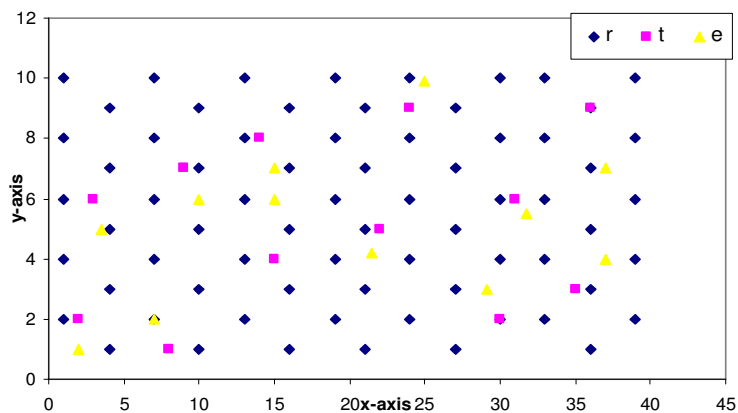

## 5.2. Testing the Efficiency of the Segmentation Process

Assigning unique features to each subarea is a critical task. As previously discussed, the segmentation process can be achieved in two ways (manual or autonomous). In this section, the ability of our proposed system to divide the whole tracking area into subareas is evaluated. The proposed system was tested in two different test-beds.

The autonomous segmentation process works efficiently in open environments, as in Test-bed 1. Figure 8 depicts the RSS values from each beacon node in each subarea. However, deploying an autonomous process in complex environments where obstacles and walls are situated is a challenging task as the RSS behaves irregularly in different areas. Thus, it was a challenging task for the autonomous segmentation to achieve the division process in complex environments. The proposed autonomous process was not able to divide Test-bed 2 into subareas; a manual segmentation was used instead. Figure 9 shows the irregular distribution of RSS values in a complex environment. Both figures show how the tracking area is divided into subareas.

**Figure 8:** Testing the efficiency of the autonomous segmentation process in Test-bed 2

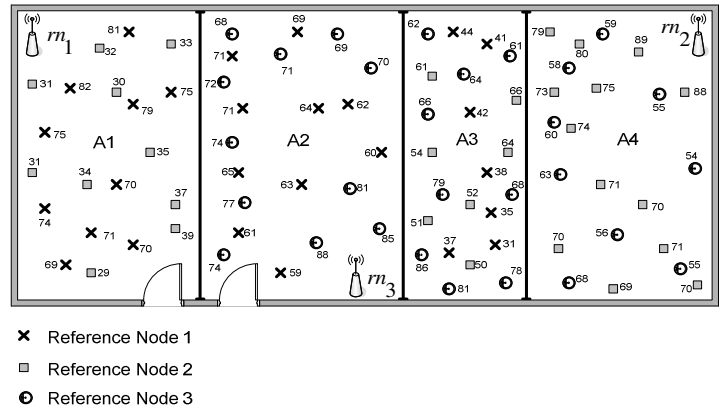

**Figure 9:** Testing the efficiency of the autonomous segmentation process in Test-bed 1

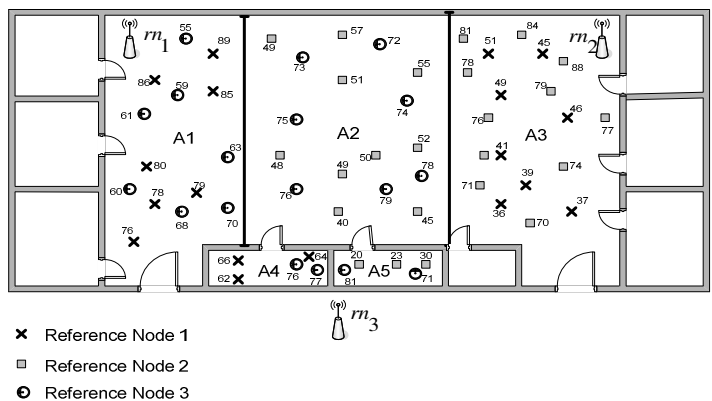

## 6. Discussion

Obliviously, the need for more reference points and measurements indicates that the offline stage is a critical task in terms of time and labor. Neural networks have been used in several fingerprinting localization approaches [20, 21, 22] in order to reduce the total number of the collected reference points, and therefore trim down the time needed to complete the offline phase. Conversely, through experiments, it has been shown that neural network-based approaches (such as RBF) performs well in the area on which it has been trained, and consequently, neural network modules need to be trained at each grid point in order to offer reasonable localization accuracy. This procedure adds time and power consumption to the whole WSN.

The approach presented in this paper does not require the collection of a large number of reference points as it's based on identifying a unique feature for each subarea. Consequently, the searching procedure can be processed in a short time period compared to [20, 21, 22].

An efficient WSN-based fingerprinting localization approach was implemented in [22]. The implementation of this approach was limited in scale as it was implemented in a small area (300 × 300 cm). Conversely, the approach proposed in this paper was deployed in two applicable areas (Test-bed 1 and Test-bed 2) with sizes of (41.5 × 11.3 m) and (30.5 × 11.3 m) respectively.

A significant localization system was proposed in [14] which offers good localization accuracy. However, it is based on the nearest neighbor method in order to compute the target's coordinates. This includes finding out the two nearest points to the estimated reference point based on the collected RSS values. However, the 3NNF first identifies the subarea where the target node is located in, and then triangulate its location using the nearest neighbor reference points. Figure 10 shows the localization error for both the 2 nearest neighbors and 3NNF methods.

**Figure 10:** The distance error based on the number of reference points (r)

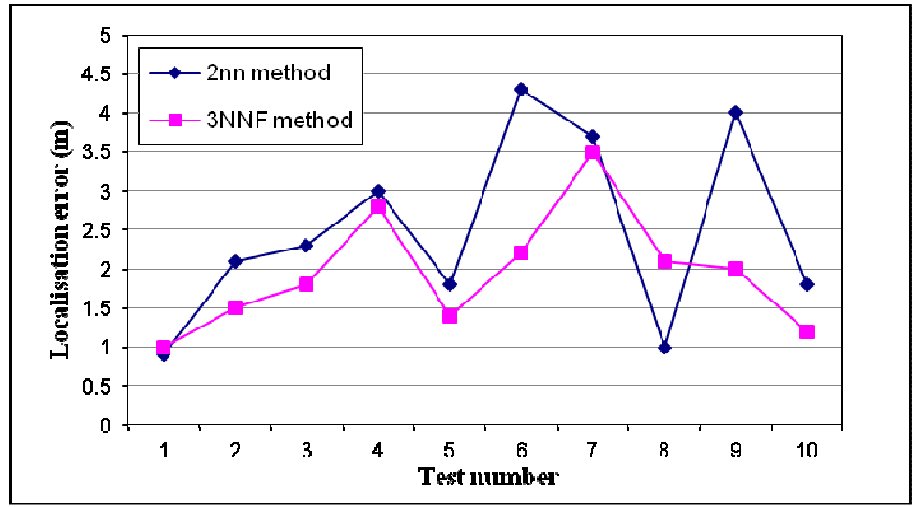

The proposed system includes dividing the whole area into subareas based on unique features which uniquely distinguish each subarea. For most previous fingerprinting systems, any changes to the environments, which alter the collected and stored features at the scenes, require recollecting the predefined data. However, the approach proposed in this paper does not require reference points to be recollected from the whole tracking area in the case of changes. It only requires data to be recollected from the subarea which has been changed. Table 3 compares the existing fingerprinting localization systems.

**Table 3:**    Evaluating the existing fingerprinting localization systems

| Localization method | Area size (meter) | Number of collected reference points | Localization error (meter) |
|---|---|---|---|
| Approach [12] | - | 132 | 1 − 4 |
| Approach [13] | Hall | 441 | 0.5 − 7.8 |
| RADAR [14] | 43.5 × 22.5 | 70 | 2 − 3 |
| Approach [17] | 23 × 17.5 | >100 | 1.78 |
| Approach [18] | - | 300 | 0.2 − 2.2 |
| Approach [19] | 25.5 × 24.5 | 200 | 2 − 3 |
| Approach [22] | 0.3 × 0.3 | 121 | 0.1 − 0.4 |
| RBF approach | 41.5 × 11.3 | 70 | 1 − 6 |
| 3NNF approach | 41.5 × 11.3 | 70 | 1 − 3.5 |

## 7.  Conclusion and Future Work

RSS-based localization systems are more competitive in terms of both accuracy and cost compared to other localization systems. Fingerprinting methods suffer from the fact that they require a large database and a long training phase; this increases the size and computational burden of the database. In this paper, three advantages were gained over the existing approaches. First, the total number of reference points that had to be collected during the offline phase was reduced. Second, the proposed approach offers good localization accuracy (1-3 meters). Third, a problem which might arise when any changes on the tracking area has been overcome by dividing the tracking area into subareas.

In this paper, the researchers have designed and implemented a segmentation process in order to divide a tracking area into small areas. The proposed autonomous method only works well in open environments and cannot be used in complicated ones. Manual segmentation can be used in any environment but it requires additional effort and time, in addition to experience regarding the system and how to divide the tracking area. One of the most important developments and improvements that should be added to research in the future is to improve the autonomous segmentation process by adopting Artificial Intelligence, in addition to increasing localization accuracy by adopting other localization methods.